\documentclass{aastex} 

\begin{document}

\title{Constraints on Inner Disk Evolution Timescales:\\ 
A Disk Census of the $\eta$ Chamaeleontis Young Cluster}

\author{Karl E. Haisch Jr.} 
\affil{Physics Department, Utah Valley State College,\\ 
800 West University Parkway, Orem, UT 84058-5999, U.S.A.}

\author{Ray Jayawardhana} 
\affil{Department of Astronomy \& Astrophysics, University of Toronto,\\ 
Toronto, Ontario M5S 3H8, CANADA}

\and

\author{Jo\~ao Alves} 
\affil{European Southern Observatory, Garching, Germany}

\begin{abstract}
We present new $L^\prime$-band (3.8$\mu$m) observations of stars in the 
nearby ($\sim$97 pc) young ($\sim$6 Myr) compact cluster around $\eta$ 
Chamaeleontis, obtained with the European Southern Observatory's Very Large 
Telescope in Paranal, Chile. Our data, combined with $J,H,K_{s}$ photometry 
from the 2-Micron All Sky Survey, reveal that only two of the 12 members surveyed 
harbor $L^\prime$-band excesses consistent with optically thick inner disks; 
both are also likely accretors. Intriguingly, two other stars with possible 
evidence for on-going accretion, albeit at very low rates, do not show 
significant infrared excess: this may imply substantial grain growth and/or 
partial clearing of the inner disk region, as expected in planet formation 
scenarios. Our findings suggest that $\eta$ Cha stars are in an epoch when 
disks are rapidly evolving, perhaps due to processes related to planet 
building, and provide further constraints on inner disk lifetimes. 

\end{abstract}

 
\section{Introduction}

There is growing evidence that circumstellar disks undergo substantial
evolution --development of inner disk cavities, grain growth, dust
settling, and decrease in accretion rates-- within the first several
million years of a star's life (e.g., Haisch, Lada, \& Lada 2001b;
Jayawardhana et al. 2001; Hogerheijde et al. 2003). Nearby young
stellar clusters constitute superb laboratories for detailed studies of
the physical processes and the timescales of such disk evolution and
planet formation. In particular, the
8-Myr-old TW Hydrae Association has proven to be a treasure trove: its
members exhibit a wide variety behavior, from classical T Tauri accreting
disks, to planetary debris systems, to systems without measureable disk
emission at $\lambda \leq 10 \mu m$ (Jayawardhana et al. 1998, 1999). 
The diverse disk properties suggest that the TW Hya stars are at an age 
when disks are rapidly evolving, through coagulation of dust and dissipation of gas.

The recently identified $\eta$ Chamaeleontis cluster (Mamajek, Lawson, 
\& Feigelson 1999) provides another superb sample in this interesting age 
range and offers the prospects of finding other disks ``in transition". The 
18 known primaries of this group, first identified by means of their X-ray 
emission and later confirmed with optical spectroscopy, include three 
early-type stars and 15 late-type (K5-M5) stars (Lyo et al. 2004; Luhman 
\& Steeghs 2004). {\it Hipparcos} measurements place the brighter stars at 
a distance of $\sim$ 97 pc. The late-type H$\alpha$ emission-line stars have 
the properties of pre-main-sequence T Tauri stars such as large Li abundance, 
high magnetic activity, and, assuming they are also at $\sim$ 97 pc, bolometric 
luminosities 1-2 mag above the main sequence. Recently, Luhman \& Steeghs (2004) 
have derived an age of $6^{+2}_{-1}$~Myr for the cluster, which is consistent 
with age estimates by Mamajek, Lawson, \& Feigelson (1999) and Lawson \& Feigelson 
(2001) within the uncertainties.

Here we report a disk census of the $\eta$ Chamaeleontis cluster in the
$L^\prime$-band, derive the inner disk fraction, compare it to other
clusters and young associations and discuss constraints on inner disk
evolution timescales.

\section{Observations}

We obtained $L^\prime$-band (3.8$\mu$m) images in service mode for 12 known late-type members of the $\eta$ Chamaeleontis cluster using the Infrared Spectrograph And Array Camera (ISAAC; Moorwood 1997) at the Very Large Telescope (VLT) UT1 during the period 2004 March 04 - 06. The observations were carried out in the long-wavelength LW4 mode of ISAAC, which is equipped with a 1024$\times$1024 InSb Aladdin array from Santa Barbara Research Center. This mode yields a pixel scale of 0.071 arcsec pixel$^{-1}$ with a field of view of 73 arcsec$^{2}$. The data were acquired in the chop-nod mode with 15\arcsec offsets in order to cancel out the variable thermal background. Each observation consisted of two sets of 9 coadded frames with integration times of 0.11 seconds. The total integration time for each star was 1.98 seconds. The $J, H$ and $K_{s}$ (1.25, 1.65, and 2.1 $\mu$m) observations of each cluster member were taken from the Two Micron All-Sky Survey (2MASS) data archive.\footnote{http://www.ipac.caltech.edu/2mass/}

All observations were reduced using the standard ISAAC reduction pipeline for long-wavelength imaging data. Flat fields for each night were obtained by subtracting three images taken at 1.0, 2.0, and 2.4 airmasses. These flat fields were normalized and divided into the sky subtracted object frames to produce the final images of each object. The standard star HD106965 was observed on the same nights and through the same range of airmass as the target stars. Zero points and extinction coefficients were established for each night. The photometric accuracy of our observations is typically $\pm$ 0.05 mag.

\section{Results}

In Table~\ref{table1} we list the $JHK_{s}L^\prime$ magnitudes, and $J$ -- $H$, $H$ -- $K_{s}$, and $K_{s}$ -- $L^\prime$ colors for all members of the $\eta$ Chamaeleontis cluster. The $JHK_{s}$ magnitudes and colors have been transformed to the CIT photometric system using 2MASS to CIT transformation equations (see previous footnote). The $L^\prime$ magnitudes were calibrated in the CIT system. Since all of the $\eta$ Chamaeleontis members have known spectral types, we have determined their intrinsic photospheric $H$ -- $K_{s}$ and $K_{s}$ -- $L^\prime$ colors, and hence the excesses above the photospheric emission; these are also included in Table~\ref{table1}. In Figure~\ref{figure1} we present the $JHK_{s}L^\prime$ color-color diagram for the $\eta$ Chamaeleontis members. Late-type stars are identified by their RECX number or ECHA number. In the figure, we plot the locus of points corresponding to the unreddened main sequence (which encompasses the range of spectral types of the cluster members) as a solid line and the locus of positions of giant stars as a heavy dashed line. The intrinsic colors of giant and dwarf stars were taken from Bessell \& Brett (1988) and transformed into the CIT system. The two leftmost parallel dashed lines define the reddening band for main sequence stars and are parallel to the reddening vector. The classical T Tauri star (CTTS) locus is plotted as a dot-dashed line (Meyer, Calvet, \& Hillenbrand 1998). The reddening law of Cohen et al. (1981), derived in the CIT system and having a slope of 2.750, has been adopted.

A very small fraction of the sources fall outside and to the right of 
the reddening lines in the infrared excess region of the $JHK_{s}L^\prime$ color-color diagram. Indeed, 2/12 (17\% $\pm$ 11\%) of the sources in the $\eta$ Chamaeleontis cluster have colors exhibiting $JHK_{s}L^\prime$ infrared excess emission indicative of circumstellar disks. This low fraction is supported by the intrinsic excesses for these stars calculated in Table ~\ref{table1}. The quoted uncertainty represents the statistical standard error (i.e., $\sqrt{[DF(1 - DF)]/N}$; DF = disk fraction) in our derived excess/disk fraction. We note that this excess fraction depends on the adopted reddening law and to a lesser extent on the photometric system used to plot the positions of the main sequence stars and the reddening bands. We have calculated the $JHK_{s}L^\prime$ infrared excess fraction for two other reddening laws obtained in different photometric systems (Koornneeef 1983; Rieke \& Lebofsky 1985), and find excess fractions identical to that derived above. This is perhaps not surprising given the absence of measured extinctions among the cluster members (Luhman \& Steeghs 2004). Additionally, we have considered the effects of binarity on the derived excess fractions, since a number of cluster members are known 
(RECX 1, RECX 7, RECX 9; K\"{o}hler \& Petr-Gotzens 2001; Lyo et al. 2003, 2004) or suspected 
(RECX 12; K\"{o}hler \& Petr-Gotzens 2001; Lawson et al. 2001; Haisch et al. 2004) to be binary systems. Examining three different sets of PMS models (Baraffe et al. 1998; Palla \& Stahler 1999; Seiss, Dufour, \& Forestini 2000), we calculate that the presence of a binary will change our photometry by $<$ 0.04 mag, within our measured photometric error. Therefore, we conclude that binarity will not affect our calculated excess (disk) fraction.

\section{Discussion}

The disk fraction we derive for the $\eta$ Chamaeleontis cluster is 
significantly smaller than that (60\% $\pm$ 13\%) found by Lyo et al. (2003). Figure~\ref{figure2} shows the $JHKL$ color-color diagram of Lyo et al. after transforming their $JHKL$ photometry to the CIT photometric system. Late-type stars are identified by their RECX number or ECHA number, and early-type stars are identified by name. 

The large difference in the derived disk fraction between our study and that of Lyo et al. is only apparent when the $L$-band data are included in the analysis. In Figure~\ref{figure3} (a), we plot the $JHK_{s}$ color-color diagram using the 2MASS photometry adopted in our study, and in Figure~\ref{figure3} (b) we plot the $JHK$ photometry of Lyo et al. As one can see, none of the $\eta$ Chamaeleontis members in the present survey show $JHK_{s}$ infrared excesses using the 2MASS colors, while 2/14 (14\% $\pm$ 9\%) show $JHK$ excesses in the Lyo survey. These results are similar to those obtained from examining the intrinsic $H$ -- $K$ excesses for each member (Table~\ref{table1}) and from published surveys of other clusters with a similar age (Haisch, Lada, \& Lada 2001b). We note that, in many cases, the $J,H$, and/or $K$-band magnitudes observed by Lyo et al. are $\sim$ 0.05 - 0.20 magnitudes fainter than the photometry in our survey, even after transforming all photometry to the same photometric system. Similarly, the Lyo et al. $L$-band magnitudes are typically 0.10 - 0.30 magnitudes brighter than our $L^\prime$ data. Thus, the $K - L$ colors in the Lyo et al. study are typically 0.15 - 0.50 magnitudes redder than those derived in our study, thereby leading to a higher calculated $JHKL$ disk fraction.

There are six additional known members of the $\eta$ Chamaeleontis cluster which are not covered in our survey. Three of these ($\eta$ Cha, RS Cha, and HD 75505) were observed at $JHKL$ by Lyo et al. (2004) and were not found to possess infrared excesses. The remaining three cluster members have no published $L$-band photometry. They do not exhibit infrared excess emission in a $JHK$ color-color diagram, however future $L$-band observations of these stars may show that they are indeed surrounded by disks. If so, this would still imply an upper limit on the disk fraction in the $\eta$ Chamaeleontis cluster of 28\% $\pm$ 13\%.

Based on H$\alpha$ emission line profiles in high-resolution optical spectra, Lawson 
et al. (2004) claimed that four of the $\eta$ Cha stars are accreting from circumstellar 
disks. One of them --ECHA J0843.3-7905-- has an H$\alpha$ equivalent width of 
$\approx$90 \AA, characteristic of classical T Tauri stars, and also exhibits a clear 
$L^\prime$ excess in our data. According to these authors, the other $L^\prime$ excess 
star in our census, RECX 11, may also be accreting, given its H$\alpha$ profile, 
though the equivalent width is only 3 \AA. We note that it is difficult to distinguish 
between weak accretion and stellar chromospheric activity, 
sometimes even with high resolution spectra (e.g. Walter \& Barry 1991; 
Jayawardhana, Mohanty \& Basri 2003). Intriguingly, two other stars --RECX 5 and 9-- 
that Lawson et al. claim to be accretors based on their H$\alpha$ profiles, 
albeit at very low rates, do not
harbor $L^\prime$ excesses consistent with optically thick disks in our
survey. This could be the result of substantial grain growth and/or
partial clearing of the inner disk region, as expected in theories of
planet formation (e.g., Bryden et al. 1999). A similar situation has been
found for two stars and one brown dwarf in the 8-Myr-old TW Hydrae association. 
TW Hya itself and Hen 3-600A appear to be accreting at low rates despite
negligible excesses at $\lesssim$5 $\mu$m (Calvet et al. 2002 and references therein). 
So does the recently identified brown dwarf 2MASS J1207-3932 in that 
association (Mohanty, Jayawardhana \& Barrado y Navascu\'es 2003; Jayawardhana
et al. 2003; Sterzik et al. 2004). If $\eta$ Cha stars are found to have 
longer-wavelength excess despite the lack of significant near-infrared excess, 
it would imply that they are at an age of rapid inner disk evolution and this cluster 
would constitute an excellent laboratory for investigating planet formation 
processes. 

Our derived disk fraction for the $\eta$ Chamaeleontis cluster, combined with similar published $JHKL$ disk fractions for young clusters (Haisch, Lada, \& Lada 2001b; Oliveira, Jeffries, \& van Loon 2004), yields an overall disk lifetime of $\sim$ 6.5 Myr. This supports and strengthens the short overall disk lifetime of $\sim$ 6 Myr found by Haisch, Lada, \& Lada (2001b). This timescale depends to some extent on the mean age of the $\eta$ Chamaeleontis cluster (the age of the NGC 2362 cluster in Figure 1 of Haisch, Lada, \& Lada (2001b) has recently been well constrained; Moitinho et al. 2001). We have adopted a mean age for the $\eta$ Chamaeleontis cluster of 6 Myr (Luhman \& Steeghs 2004). Adopting an older age of 9 Myr (Lawson \& Feigelson 2001) for the cluster would likely indicate that the decline in disk fraction with time does not follow a single linear fit. Rather, after a rapid decline during which the dust in the inner disk is dissipated or accumulates into larger bodies, the disk fraction in clusters would decrease more slowly, with a small number of stars ($\sim$ 10\%) retaining their disks for times comparable to the cluster age, as is observed in the TW Hya association (e.g., Jayawardhana et al. 1999).

The $L^\prime$-band emission from circumstellar disks is produced very close ($\leq$ 0.1 AU) to the stellar surface. However, some millimeter studies of stars in young clusters suggest that the rapid decline in the circumstellar disk fraction with cluster age observed in the inner disk regions may also apply to the outer disk regions ($>$ 1 AU) where most planet formation is likely to occur (Haisch, Lada, \& Lada 2001a, Carpenter 2002, Lada \& Haisch 2005). This places important constraints on the timescale allowed for building gas giant planets around such stars.
 
\acknowledgements

Based on observations collected at the European Southern Observatory, Chile (ESO Programme 072.C-0364(A). We thank the ESO VLT staff for their outstanding support. 2MASS is a joint project of the University of Massachusetts and the Infrared Processing and Analysis Center/California Institute of Technology, funded by the National Aeronautics and Space Administration and the National Science Foundation. This work was supported in part by an NSERC grant to R.J. 


\newpage

\clearpage
\begin{deluxetable}{ccccccccccc}
\rotate
\footnotesize
\tablecaption{JHK$_{s}$L$^\prime$ Magnitudes and Colors \label{table1}}
\tablewidth{0pt}
\tablehead{Star & $J$\tablenotemark{a} & $H$\tablenotemark{a} &
$K_{s}$\tablenotemark{a} & $L^\prime$ & $(J-H)$ & $(H-K_{s})$ &
$(K_{s}-L^\prime)$ & $(H - K_{s})_{ex}$\tablenotemark{b} & $(K_{s}-L^\prime)_{ex}$\tablenotemark{b} & ST\tablenotemark{c}}
\startdata
 RECX 1 &  8.15 &  7.50 &  7.37 &  7.18 & 0.65 & 0.13 & 0.19 & 0.02 & 0.07 & K6\cr
 RECX 3 & 10.34 &  9.65 &  9.45 &  9.31 & 0.69 & 0.20 & 0.14 & -0.06 & -0.19 & M3.25\cr
 RECX 4 &  9.52 &  8.78 &  8.65 &  8.45 & 0.74 & 0.13 & 0.20 & -0.08 & -0.03 & M1.75\cr
 RECX 5 & 10.77 & 10.10 &  9.89 &  9.57 & 0.67 & 0.21 & 0.32 & -0.07 & -0.05 & M4\cr
 RECX 6 & 10.22 &  9.58 &  9.32 &  9.13 & 0.64 & 0.26 & 0.19 & 0.01 & -0.13 & M3\cr
 RECX 7 &  8.41 &  7.76 &  7.67 &  7.45 & 0.65 & 0.09 & 0.22 & -0.03 & 0.10 & K6\cr
 RECX 9 & 10.26 &  9.67 &  9.37 &  8.98 & 0.59 & 0.30 & 0.39 & 0.00 & -0.01 & M4.5\cr
RECX 10 &  9.63 &  8.92 &  8.76 &  8.48 & 0.71 & 0.16 & 0.28 & -0.04 & 0.07 & M1\cr
RECX 11 &  8.73 &  8.04 &  7.69 &  7.03 & 0.69 & 0.35 & 0.66 & 0.24 & 0.55 & K5.5\cr
RECX 12 &  9.31 &  8.68 &  8.44 &  8.05 & 0.63 & 0.24 & 0.39 & -0.01 & 0.06 & M3.25\cr
eChaJ0841.5-7853 & 11.81 & 11.24 & 11.01 & 10.63 & 0.57 & 0.23 & 0.38 & -0.08 & -0.02 & M4.75\cr
eChaJ0843.3-7905 & 10.50 &  9.83 &  9.46 &  8.14 & 0.67 & 0.37 & 1.32 & 0.12 & 0.99 & M3.25\cr
\enddata
\tablenotetext{a}{$JHK_{s}$ magnitudes are taken from the 2MASS database and transformed to the CIT photometric system.}
\tablenotetext{b}{Intrinsic infrared excess based on 2MASS photometry and spectral type.}
\tablenotetext{c}{Spectral Types taken from Luhman \& Steeghs (2004)}
\end{deluxetable}

\clearpage

\figcaption[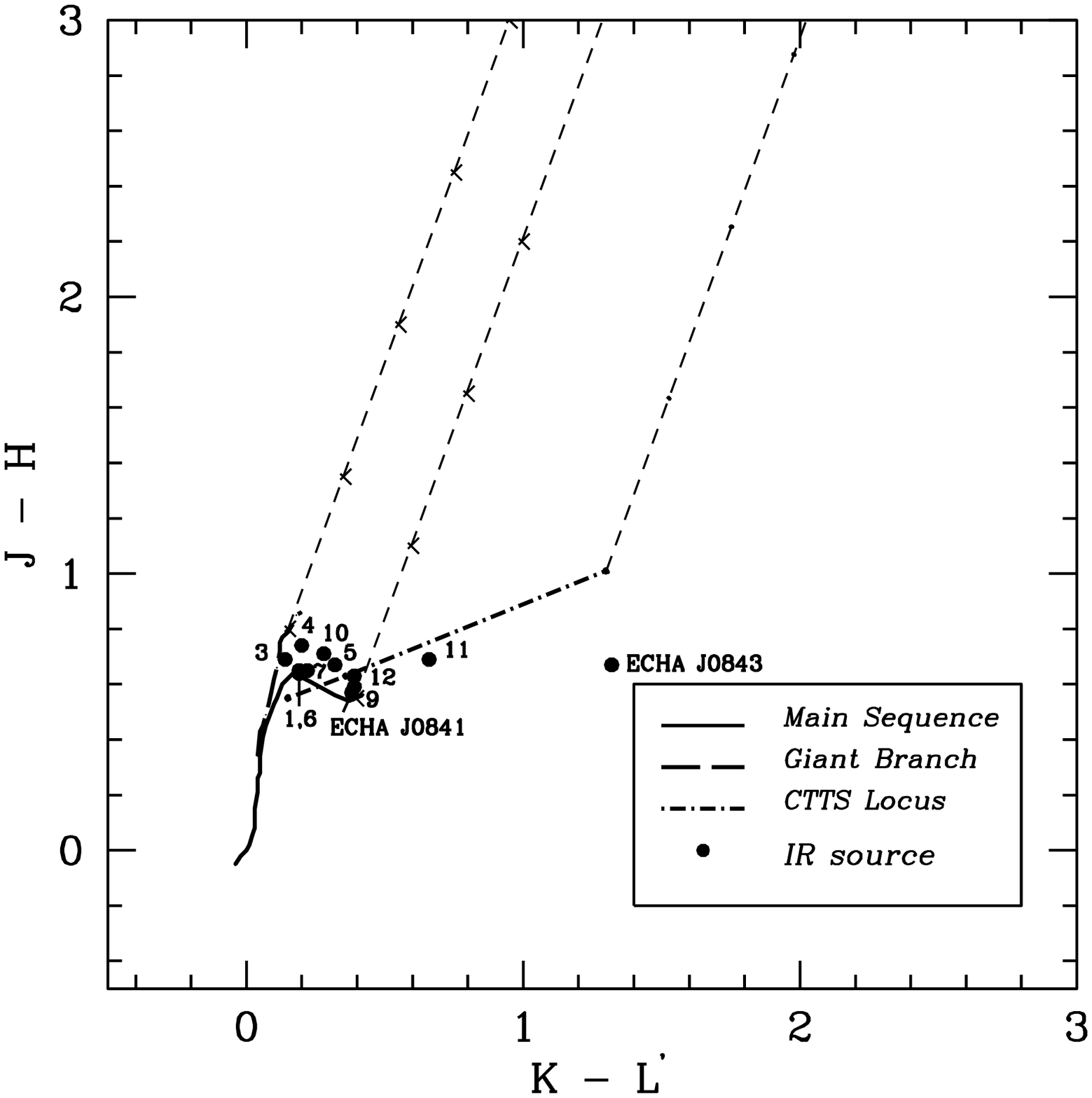]
{$JHK_{s}L^\prime$ color-color diagram for the $\eta$ Chamaeleontis cluster members. In the diagram, the locus of points corresponding to the unreddened main sequence is plotted as a solid line, the locus of positions of giant stars is shown as a heavy dashed line and the CTTS locus as a dot-dashed line. The intrinsic colors of giant and dwarf stars were taken from Bessell \& Brett (1988) and transformed into the CIT system. The two leftmost dashed lines define the reddening band for main sequence stars and are parallel to the reddening vector, the magnitude and slope of which was taken from Cohen et al. (1981) for the CIT system. The rightmost dashed line is parallel to the reddening band. Late-type stars are identified by their RECX number or ECHA number.
\label{figure1}
}
\figcaption[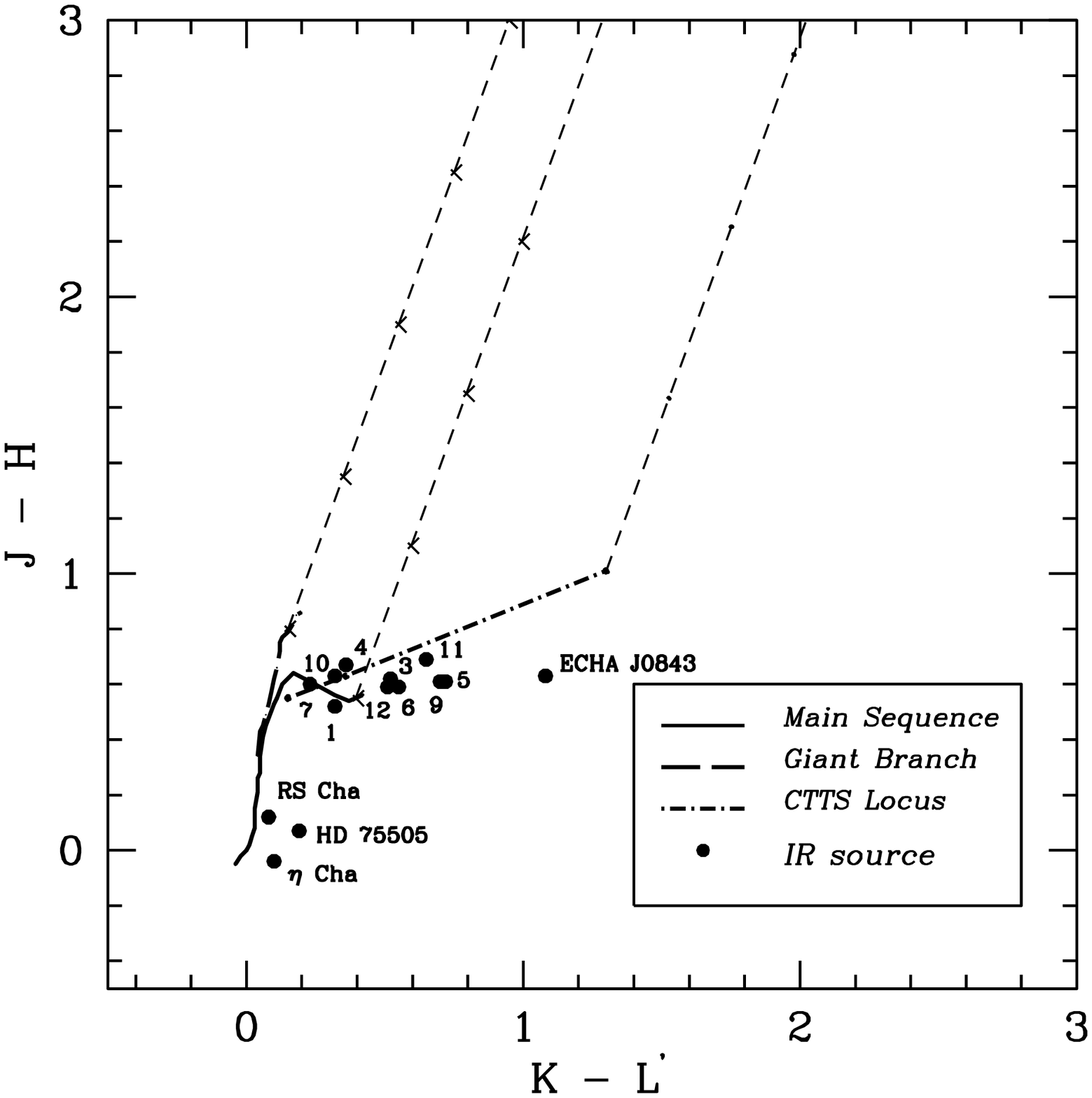]
{$JHK_{s}L^\prime$ color-color diagram for the $\eta$ Chamaeleontis cluster members observed by Lyo et al. (2003) transformed to the CIT photometric system. In the diagram, the locus of points corresponding to the unreddened main sequence is plotted as a solid line, the locus of positions of giant stars is shown as a heavy dashed line and the CTTS locus as a dot-dashed line. The intrinsic colors of giant and dwarf stars were taken from Bessell \& Brett (1988) and transformed into the CIT system. The two leftmost dashed lines define the reddening band for main sequence stars and are parallel to the reddening vector, the magnitude and slope of which was taken from Cohen et al. (1981) for the CIT system. The rightmost dashed line is parallel to the reddening band. Late-type stars are identified by their RECX number or ECHA number, and early-type stars are identified by name.
\label{figure2}
}
\figcaption[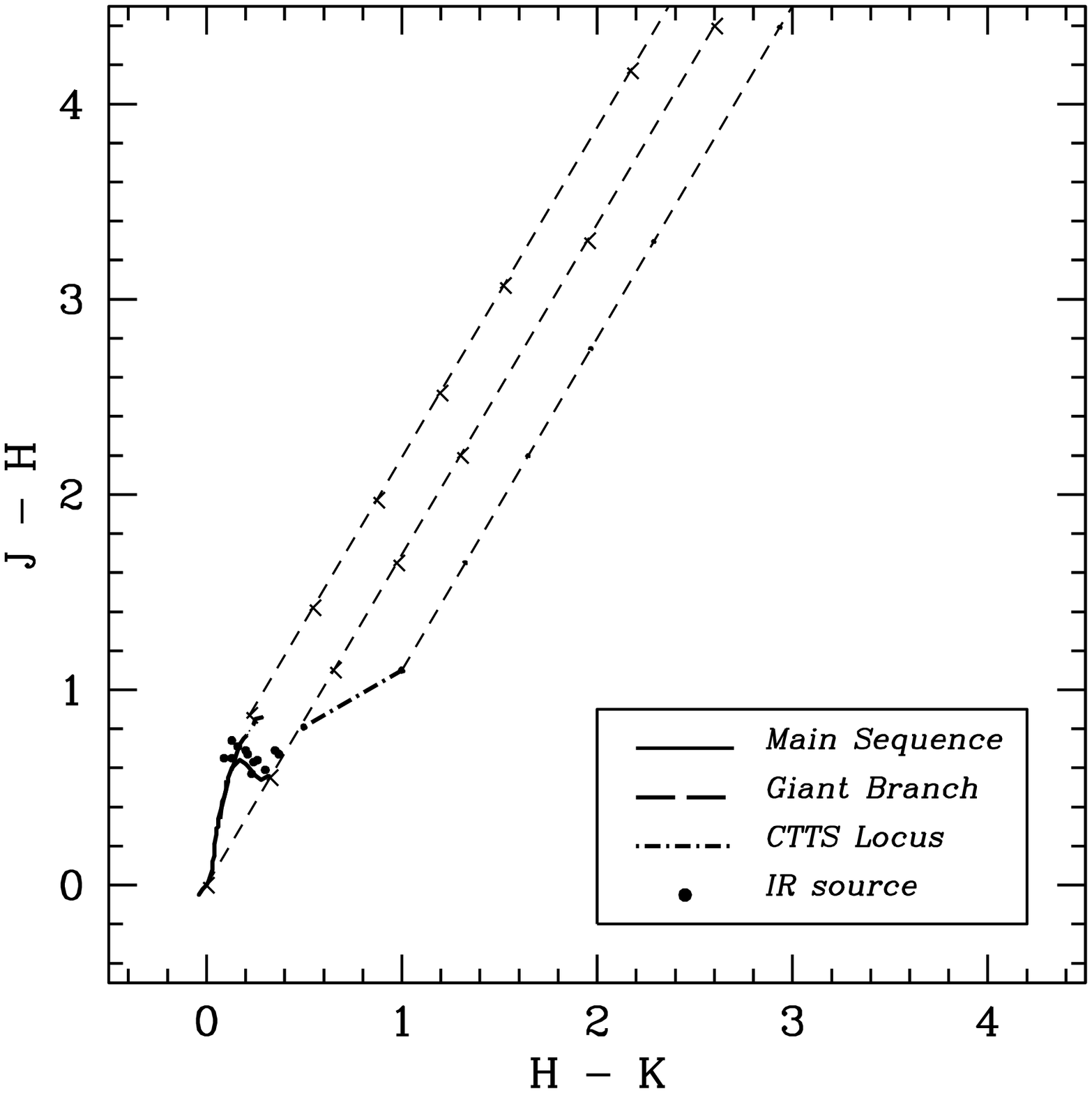]
{(a) $JHK_{s}$ and (b) $JHK$ color-color diagrams for the $\eta$ Chamaeleontis cluster members from the 2MASS catalog and those observed by Lyo et al. (2003) respectively, transformed to the CIT photometric system. In the diagram, the locus of points corresponding to the unreddened main sequence is plotted as a solid line, the locus of positions of giant stars is shown as a heavy dashed line and the CTTS locus as a dot-dashed line. The intrinsic colors of giant and dwarf stars were taken from Bessell \& Brett (1988) and transformed into the CIT system. The two leftmost dashed lines define the reddening band for main sequence stars and are parallel to the reddening vector, the magnitude and slope of which was taken from Cohen et al. (1981) for the CIT system. The rightmost dashed line is parallel to the reddening band.
\label{figure3}
}


\begin{thebibliography}{}

\bibitem[Baraffe et al.\ 1998]{bcah98} Baraffe, I., Chabrier, G., Allard, F., \& Hauschildt, P. H. 1998, \aap, 337, 403
\bibitem[Bessell \& Brett 1988]{bb88} Bessell, M. S. \& Brett, J. M. 1988, \pasp, 100, 1134
\bibitem[Bryden et al.\ 1999]{bryden99}Bryden, G., Chen, X., Lin, D.N.C., Nelson, R.P., \& Papaloizou, J.C.B. 1999, ApJ, 514, 344
\bibitem[Calvet et al. 2002]{cal02} Calvet, N., D'Alessio, P., Hartmann, L., Wilner, D., Walsh, A., \& Sitko, M. 2002, ApJ, 568, 1008
\bibitem[Carpenter 2002]{car02} Carpenter, J. M. 2002, \aj, 124, 1593
\bibitem[Cohen et al.\ 1981]{coh81} Cohen, J. G., Frogel, J. A., Persson, S. E., \& Elias, J. H. 1981, \apj, 249, 481
\bibitem[Haisch et al.\ 2005]{hai05} Haisch, Jr., K. E., Jayawardhana, R., Brandeker, A., \& Mardones, D. 2005, in Science with Adaptive Optics, Proceedings of the ESO Workshop held in Garching, Germany, 16 - 19 September 2003, eds. W. Brandner \& M. Kasper, Springer-Verlag, 177
\bibitem[Haisch, Lada, \& Lada 2001b]{hll01b} Haisch, Jr., K. E., Lada, E. A., \& Lada, C. J. 2001b, \apj, 553, L153
\bibitem[Haisch, Lada, \& Lada 2001a]{hll01a} Haisch, Jr., K. E., Lada, E. A., \& Lada, C. J. 2001a, \aj, 121, 2065
\bibitem[Hogerheijde et al.\ 2003]{hog03} Hogerheijde, M. R., Johnstone, D., Matsuyama, I., Jayawardhana, R., \& Muzerolle, J. 2003, \apj, 593, 101
\bibitem[Jayawardhana, Mohanty, \& Basri 2003]{jmb03} Jayawardhana, R., Mohanty, S., \& Basri, G. 2003, \apj, 592, 282
\bibitem[Jayawardhana et al. 2003]{j03} Jayawardhana, R., Ardila, D.R., Stelzer, B., \& Haisch, K.E. 2003, AJ, 126, 1515 
\bibitem[Jayawardhana et al.\ 2001]{jay01} Jayawardhana, R., Wolk, S. J., Barrado y Navascu\'{e}s, D., Telesco, C. M., \& Hearty, T. J. 2001, \apj, 550, 197
\bibitem[Jayawardhana et al.\ 1999]{jay99} Jayawardhana, R., Hartmann, L., Fazio, G., Fisher, R. S., Telesco, C. M., \& Pi\~{n}a, R. K. 1999, \apj, 521, L129
\bibitem[Jayawardhana et al.\ 1998]{jay98} Jayawardhana, R., Fisher, S., Hartmann, L., Telesco, C., Pina, R., \& Fazio, G. 1998, \apj, 503, 79
\bibitem[K\"{o}hler \& Petr-Gotzens 2001]{koh01} K\"{o}hler, R., \& Petr-Gotzens, M. G. 2001, \aj, 124, 2899
\bibitem[Koornneef 1983]{kor83} Koornneef, J. 1983, \aap, 128, 84
\bibitem[Lada \& Haisch 2005]{lh05} Lada, E. A., \& Haisch, Jr., K. E. 2005, \aj,in press
\bibitem[Lawson \& Feigelson 2001]{lf01} Lawson, W. A., \& Feigelson, E. D. 2001, in From Darkness To Light: Origin and Evolution of Young Stellar Clusters, ASP Conf. Ser. 243, eds. T. Montmerle and P. Andr\'{e}, San Francisco: ASP, p. 591
\bibitem[Lawson et al.\ 2001]{law01} Lawson, W. A., Crause, L. A., Mamajek, E. E., \& Feigelson, E. D. 2001, \mnras, 321, 57
\bibitem[Lawson et al.\ 2004]{law04} Lawson, W. A., Lyo, A.-Ran, \& Muzzerolle, J. 2004, MNRAS, 351, L39
\bibitem[Luhman \& Steeghs 2004]{luh04} Luhman, K. L., \& Steeghs, D. 2004, \apj, 609, 917
\bibitem[Lyo et al.\ 2004]{lyo04} Lyo, A.-Ran, Lawson, W. A., Feigelson, E. D., \& Crause, L. A. 2004, \mnras, 347, 246
\bibitem[Lyo et al.\ 2003]{lyo03} Lyo, A.-Ran, Lawson, W. A., Mamajek, E. E., Feigelson, E. D., Sung, E-C., \& Crause, L. A. 2003, \mnras, 338, 616
\bibitem[Mamajek, Lawson \& Feigelson 1999]{mlf99} Mamajek, E. E., Lawson, W. A., \& Feigelson, E. D. 1999, \apj, 516, 77
\bibitem[Meyer, Calvet, \& Hillenbrand 1997]{mch97} Meyer, M. R., Calvet, N., \& Hillenbrand, L. A. 1997, \aj, 114, 233
\bibitem[Mohanty et al. 2003]{mjb03} Mohanty, S., Jayawardhana, R., \& Barrado y Navascu\'es, D. 2003, ApJ, 593, L109
\bibitem[Moitinho et al.\ 2001]{moi01} Moitinho, A., Alves, J., Hu\'{e}lamo, N., \& Lada, C. J. 2001, \apj, 563, 73
\bibitem[Moorwood 1997]{mor97} Moorwood, A. 1997, SPIE, 2871, 1146
\bibitem[Oliveira, Jeffries, \& van Loon 2004]{ojl04} Oliveira, J. M., Jeffries, R. D., \& van Loon, J. Th. 2004, \mnras, 347, 1327
\bibitem[Palla \& Stahler 1999]{ps99} Palla, F., \& Stahler, S. W. 1999, \apj, 525, 772
\bibitem[Rieke \& Lebofsky 1985]{rl85} Rieke, G. H., \& Lebofsky, M. J. 1985, \apj, 288, 618
\bibitem[Seiss, Dufour, \& Forestini 2000]{sdf00} Seiss, L., Dufour, E., \& Forestini, M. 2000, \aap, 358, 593
\bibitem[Sterzik et al. 2004]{s04} Sterzik, M.F., et al. 2004, A\&A, 427, 245
\bibitem[Walter \& Barry 1991]{wb91} Walter, F. M., \& Barry, D. C. 1991, in The Sun in Time, ed. C. P. Sonett, M. S. Giampapa, \& M. S. Matthews (Tucson: Univ. Arizona Press), 633

\end{thebibliography}
\end{document}